# Bias of determinacy coefficients in confirmatory factor analysis based on categorical variables

André Beauducel[*], Norbert Hilger, & Anneke Weide

*University of Bonn, Department of Psychology, Germany*

## Abstract

The relevance of determinacy coefficients as indicators for the validity of factor score predictors has regularly been emphasized. Previous simulation studies revealed biased determinacy coefficients for factor score predictors based on categorical variables. Therefore, and because there are different possibilities to compute determinacy coefficients, the present study compared bias of determinacy coefficients for the best linear factor score predictor and for a correlation-preserving factor score predictor based on confirmatory factor models with observed variables with 2, 4, 6, and 8 categories and maximum likelihood estimation, diagonally weighted least squares estimation, and Bayesian estimation. Positive bias was found when data were based on variables with two categories, population factors were correlated, and when there were unmodeled cross-loadings. Based on the results, the correction for sampling error, the use of maximum likelihood or Bayesian parameters, and data with at least four categories are recommended to avoid overestimation of parameter-based determinacy coefficients.

[*]Address for correspondence: Dr. André Beauducel, University of Bonn, Department of Psychology, Kaiser-Karl-Ring 9, 53111 Bonn, Germany, email: beauducel@uni-bonn.de

In factor analysis, we can distinguish between factors, factor scores, and factor score predictors. Factor scores refer to the values of each individual on the latent common factors. Factor scores of an individual cannot unambiguously be determined from the parameters of the factor model (Nicewander, 2019; Guttman, 1955). Nevertheless, the magnitude of individual factor scores is of interest in applied settings when decisions are made at the level of the individual. Therefore, factor score predictors have been proposed (Thurstone, 1935; Grice, 2001). As factor scores and factor score predictors are not identical, determinacy, that is, the correlation of factor score predictors with factor scores is of interest as an indicator for the validity of factor score predictors (Ferrando & Lorenzo-Seva, 2018). Fortunately, determinacy can be computed from the model parameters even when the factor scores themselves are unknown. It has been recommended to report the determinacy coefficient whenever factor score predictors are computed (Grice, 2001).

However, the determinacy coefficient has been found to be biased in exploratory factor analysis (Beauducel & Hilger, 2017a) and confirmatory factor analysis (CFA), where it may be related to the bias model parameter estimates (Beauducel & Hilger, 2017b). Diagonally weighted least squares, mean- and variance-adjusted (WLSMV) estimation has been proposed for estimation of parameters in structural equation modeling and CFA based on categorical observed variables (Muthén, du Toit, & Spisic, 1997). Simulation studies demonstrated that WLSMV is useful, especially when the number of categories of the variables is small (Li, 2016; Rhemtulla, Brosseau-Liard, & Savalei, 2012). WLSMV provides estimates compensating for the effects of categorical observed variables, so that the estimates are performed for hypothesized, normally distributed continuous variables. However, when factor score predictors are computed, the observed categorical variables, not the hypothesized continuous variables, are weighted and aggregated. As Beauducel and Hilger (2017b) noted, this implies that the categorical properties of the observed variables subsist in the factor score predictors computed from categorical variables even when the model parameter estimates are based on hypothesized continuous variables. Therefore, they proposed a determinacy coefficient for categorical variables which combines WLSMV-estimates with ML-estimates in order to prevent overestimation of determinacy. Since this WLSMV/ML determinacy coefficient for categorical variables has until now only been investigated for one-factor models, further investigation of this determinacy coefficient is necessary.

CFA based on Bayes estimation (BA) has been introduced as a promising estimation method (Muthén & Asparouhov, 2012). BA-estimation does not require to fix loadings to a single value. For example, exact zero loadings may be replaced with more realistic, approximate zeros based on informative, small-variance priors. Even when sensitivity analysis for priors has been recommended (Depaoli & van de Schoot, 2017), the use of correct zero priors with small variance in BA-estimation results in loading estimates that are close to the population loadings (Xiao, Liu & Hau, 2019). The use of plausible values has been proposed in the context of BA-estimation (Asparouhov & Muthén, 2010). Moreover, it has been shown that mean plausible values computed from BA-estimation have the same determinacies as the best linear factor score predictor computed from BA-parameters (Beauducel & Hilger, 2022). This study also shows that Budescu's (1982) correction for sampling error reduces positive bias of determinacy coefficients. However, this simulation study was limited to continuous variables and BA-estimation, so that it is unknown whether Budescu's correction works for ML-estimation and for categorical variables as well.

The effect of different conditions (population model, sample size, and number of categories) on determinacy coefficients based on parameter estimates from ML, combined WLSMV/ML, and BA-estimation has not yet been investigated. The aim of the present study was therefore to investigate the bias of determinacy coefficients for these estimation methods for different numbers of factors, different population models, and different numbers of categories of observed variables.

Whereas the conventional determinacy coefficient is an estimate for the correlation of the best linear predictor with the factor score other factor score predictors could also be of interest. For example, correlation-preserving factor score predictors (Grice, 2001; McDonald, 1981; Beauducel, Hilger, & Kuhl, in press) could be of interest when the scores are used as predictors of external criterion variables or when factor score predictors are inserted into higher order factor analysis. Beauducel et al. (in press) have shown that for several CFA models, the correlation-preserving factor score predictor can have a determinacy coefficient of a similar size as the determinacy coefficient of the best linear predictor. However, their simulation study was restricted to maximum likelihood CFA, and they did not investigate the effect of categorical variables on determinacy coefficients. Accordingly, it is not yet known whether the conditions for bias of the conventional determinacy coefficient, like the number of categories of the observed variables affect the determinacy coefficient of the correlation preserving factor score predictor in a similar way. The determinacy coefficient for the correlation preserving factor score predictor was therefore also investigated in the present study.

Although the individual factor scores are unknown in empirical research, they are known in simulation studies because they can *a priori* be defined in the context of data generation. Therefore, it is possible to compute the correlation of the *a priori* defined factor scores with the factor score predictors of the corresponding factor as a score-based determinacy. This correlation of the factor scores with the factor score predictors can be compared with the determinacy coefficient computed from the estimated model parameters. The difference between the model parameter-based determinacy coefficient and the score-based determinacy coefficient indicates how exactly the correlation of the factor score predictors with the factor scores can be approximated by the parameter-based determinacy coefficient. Thus, the difference between the parameter-based determinacy coefficient and the score-based determinacy coefficient is a measure of bias. The parameter-based determinacy coefficients should be a proxy of the score-based determinacy coefficient because the latter cannot be computed in empirical studies.

To sum up, the aim of the present study was to investigate bias of determinacy coefficients based on parameters from ML-, WLSMV/ML-, and BA-estimation, with and without correction for sampling error. This was examined for different population factor models, model misfit, number of categories, and sample sizes. As the type of factor score predictor may also affect the bias of parameter-based determinacy coefficients, the investigation was performed for the best linear factor score predictor and the correlation preserving factor score predictor. As simulation studies allow for the computation of the correlation of factor score predictors with *a priori* defined factor scores as score-based determinacy coefficients, these reference values for determinacy were compared with the parameter-based determinacy coefficients in order to estimate the amount of bias.

First, definitions of the factor model and the definitions of the determinacy coefficients investigated in the present study are presented. Second, the conditions, that is, the independent variables, and the different determinacy coefficients and their bias, that is, the dependent variables of this study are described. Finally, the data generation and results are presented and discussed.

**Definitions**

According to the factor model, the $p$ observed variables $\mathbf{x}$ can be decomposed into

$$\mathbf{x} = \mathbf{x}_M + \mathbf{x}_R = \hat{\mathbf{\Lambda}}\boldsymbol{\xi}_u + \hat{\mathbf{\Psi}}\boldsymbol{\varepsilon}_u + \mathbf{x}_R, \qquad (1)$$

where $\mathbf{x}_M$ is explained by the model and $\mathbf{x}_R$ is the residual part, $\hat{\mathbf{\Lambda}}$ is the matrix of estimated common factor loadings, $\hat{\mathbf{\Psi}}$ is the positive definite matrix of unique factor loadings, $\boldsymbol{\xi}_u$ are $q$ common factors, with $E(\boldsymbol{\xi}_u) = \mathbf{0}$, $E(\boldsymbol{\xi}_u\boldsymbol{\xi}_u') = \hat{\mathbf{\Phi}}$, $diag(\hat{\mathbf{\Phi}}) = \mathbf{I}$, and $\boldsymbol{\varepsilon}_u$ are the $p$ unique factors with $E(\boldsymbol{\varepsilon}_u) = \mathbf{0}$ and $E(\boldsymbol{\varepsilon}_u\boldsymbol{\varepsilon}_u') = \mathbf{I}$. The suffix "$u$" indicates that the common and unique factors are unavailable in empirical studies. It is furthermore expected that $E(\boldsymbol{\xi}_u\boldsymbol{\varepsilon}_u') = \mathbf{0}$, $E(\boldsymbol{\xi}_u\mathbf{x}_R') = \mathbf{0}$, and $E(\boldsymbol{\varepsilon}_u\mathbf{x}_R') = \mathbf{0}$. The covariance of observed variables $\mathbf{S}$ can be decomposed into

$$\mathbf{S} = \hat{\mathbf{\Sigma}}_M + \mathbf{S}_R = \hat{\mathbf{\Lambda}}\hat{\mathbf{\Phi}}\hat{\mathbf{\Lambda}}' + \hat{\mathbf{\Psi}}^2 + \mathbf{x}_R\mathbf{x}_R', \qquad (2)$$

where $\hat{\mathbf{\Sigma}}_M$ is the estimate of the population covariance matrix of observed variables $\mathbf{\Sigma}$ from the model parameters and $\mathbf{S}_R$ is the residual covariance matrix. Note that $\hat{\mathbf{\Psi}}^2$ may contain correlated errors, i.e., non-zero non-diagonal elements. The correlated error terms are part of the model, they will be part of $\hat{\mathbf{\Sigma}}_M$, which is of relevance for the computation of factor score predictors. When there are correlations in the empirical data that are not part of the model, they will be part of $\mathbf{x}_R\mathbf{x}_R'$ and therefore not be included into $\hat{\mathbf{\Sigma}}_M$, but they will be part of $\mathbf{S}$. When model misfit occurs, the complete covariance of observed variables is not explained by the factor model, so that $\hat{\mathbf{\Sigma}}_M \neq \mathbf{S}$.

Based on the factor model, Thurstone's best linear factor score predictor can be computed as

$$\hat{\boldsymbol{\xi}}_{BL} = \hat{\mathbf{\Phi}}\hat{\mathbf{\Lambda}}'\hat{\mathbf{\Sigma}}^{-1}\mathbf{x}. \qquad (3)$$

McDonald (1974) and Mulaik and McDonald (1978) proposed to compute the determinacy coefficient as

$$diag(Cor(\hat{\xi}_{BL},\xi_u)) \approx \mathbf{P}_{BL} = diag(\hat{\mathbf{\Phi}}\hat{\mathbf{\Lambda}}'\hat{\mathbf{\Sigma}}^{-1}\hat{\mathbf{\Lambda}}\hat{\mathbf{\Phi}})^{-1/2} diag(\hat{\mathbf{\Phi}}\hat{\mathbf{\Lambda}}'\hat{\mathbf{\Sigma}}^{-1}\hat{\mathbf{\Lambda}}\hat{\mathbf{\Phi}}) = (\hat{\mathbf{\Phi}}\hat{\mathbf{\Lambda}}'\hat{\mathbf{\Sigma}}^{-1}\hat{\mathbf{\Lambda}}\hat{\mathbf{\Phi}})^{1/2}, \quad (4)$$

which is based on matrices resulting from ML-estimation. The right side of Equation 4 denotes the parameter-based determinacy coefficient, which should be a proxy of the score-based determinacy coefficient given at the left hand side. Beauducel and Hilger (2022) found that this determinacy coefficient can also be used for mean plausible values resulting from BA-estimation. Note that it is necessary to use the sample of observed scores $\mathbf{x}$ in Equation 3 as it is impossible to compute $\mathbf{x}_M$ because of factor score indeterminacy. Accordingly, Gorsuch (1983), Grice (2001), and Heermann (1963) proposed to compute

$$diag(Cor(\hat{\xi}_{BL},\xi_u)) \approx \mathbf{P}_{SBL} = diag(\hat{\mathbf{\Phi}}\hat{\mathbf{\Lambda}}'\mathbf{S}^{-1}\hat{\mathbf{\Lambda}}\hat{\mathbf{\Phi}})^{-1/2} diag(\hat{\mathbf{\Phi}}\hat{\mathbf{\Lambda}}'\mathbf{S}^{-1}\hat{\mathbf{\Lambda}}\hat{\mathbf{\Phi}}) = (\hat{\mathbf{\Phi}}\hat{\mathbf{\Lambda}}'\mathbf{S}^{-1}\hat{\mathbf{\Lambda}}\hat{\mathbf{\Phi}})^{1/2}. \quad (5)$$

However, $\mathbf{P}_{SBL}$, that is, the determinacy coefficient based on the sample covariance matrix has already been shown to be more biased than $\mathbf{P}_{BL}$ (Beauducel, 2011), so that this coefficient was not further investigated.

In order to account for categorical variables, Beauducel and Hilger (2017b, p. 3421, Eq. 11) proposed that

$$diag(Cor(\hat{\xi}_{BLc},\xi_u)) \approx \mathbf{P}_{BLc} = diag(\hat{\mathbf{\Phi}}_c\hat{\mathbf{\Lambda}}'_c\hat{\mathbf{\Sigma}}_c^{-1}\hat{\mathbf{\Sigma}}\hat{\mathbf{\Sigma}}_c^{-1}\hat{\mathbf{\Lambda}}_c\hat{\mathbf{\Phi}}_c)^{-1/2} diag(\hat{\mathbf{\Phi}}_c\hat{\mathbf{\Lambda}}'_c\hat{\mathbf{\Sigma}}_c^{-1}\hat{\mathbf{\Lambda}}\hat{\mathbf{\Phi}}), \quad (6)$$

where "*c*" indexes the matrices resulting from WLSMV-estimation, whereas the other matrices are estimated from ML-estimation.

A correlation preserving factor score predictor discussed in ten Berge, Krijnen, Wansbeek and Shapiro (1999), Grice (2001), Beauducel and Hilger (2022) is

$$\hat{\xi}_{CP} = \hat{\mathbf{\Phi}}^{1/2}(\hat{\mathbf{\Lambda}}'\hat{\mathbf{\Sigma}}^{-1}\hat{\mathbf{\Lambda}})^{-1/2}\hat{\mathbf{\Lambda}}'\hat{\mathbf{\Sigma}}^{-1}\mathbf{x}, \quad (7)$$

with an orthogonal version (for $\hat{\mathbf{\Phi}} = \mathbf{I}$) proposed by Takeuchi, Yanai and Mukherjee (1982), which has the determinacy coefficient

$$diag(Cor(\hat{\xi}_{CP},\xi_u)) \approx \mathbf{P}_{CP} = \hat{\mathbf{\Phi}}^{1/2}(\hat{\mathbf{\Lambda}}'\hat{\mathbf{\Sigma}}^{-1}\hat{\mathbf{\Lambda}})^{-1/2}\hat{\mathbf{\Lambda}}'\hat{\mathbf{\Sigma}}^{-1}\hat{\mathbf{\Lambda}}\hat{\mathbf{\Phi}}. \quad (8)$$

The determinacy coefficients defined in Equations 4, 6, and 8 will be corrected for over-estimation $\mathbf{\Omega}$ due to sampling error by means of

$$\mathbf{\Omega}(\mathbf{P}^2) = \left(\frac{p-2}{n-p-1}\right)(1-\mathbf{P}^2) + \left(\frac{2(n-3)}{(n-p)^2-1}\right)(1-\mathbf{P}^2)^2, \quad (9)$$

where *n* represents sample size. According to Budescu (1982, p. 973, Eq. 3), the corrected determinacy coefficients are computed by $\mathbf{P}^c = (\mathbf{P^2} - \Omega(\mathbf{P^2}))^{1/2}$.

## Simulation study on bias of determinacy coefficients

*Simulation for finite populations*

The simulation study was devoted to the investigation of the magnitude and bias of model parameter-based determinacy coefficients as estimates for score-based determinacy coefficients. The populations were based on *q* = 3 factors and *p/q* = 5 variables with salient loadings on each factor. The conditions investigated in the simulation study for populations were the salient loading size $sl \in \{.40, .80\}$, no cross-loadings *cl* = 0 or one cross-loading per five variables with salient loadings of size *sl*/2 (.20 or .40), that is, *cl* = 1 (see Table 1), factor inter-correlations $\phi \in \{.00, .30\}$, sample size $n \in \{300, 900\}$, and number of categories of observed variables $c \in \{2, 4, 6, 8\}$. The thresholds for categorization of continuous normally distributed variables were stable and symmetric. In total, this resulted in 2 (*sl*) × 2 (*cl*) × 2 ($\phi$) × 4 (*c*) = 32 simulation conditions. For each condition, a population of 1,000,000 cases was drawn from multivariate normal distributions and transformed into symmetric binomial distributions with two categories. The simulated population data was then submitted to ML-estimation for continuous variables, WLSMV-estimation for categorical variables, and BA-estimation for continuous variables. All salient loadings were freely estimated. The condition with cross-loadings corresponds to a condition of model misfit because all cross-loadings were fixed to zero in model specification of ML- and WLSMV models, and the prior of the cross-loadings was fixed to zero in BA models. Factor variance were fixed to one and factor inter-correlations were fixed to zero for *r* = .00. Note that the cross-loading condition was introduced to introduce model misfit. Therefore, with ML- and WLSMV-estimation all non-salient loadings were fixed to zero (i.e., independent-cluster models were specified), so that model misfit was present in the non-zero cross-loading condition, whereas there was no model misfit when there were no cross-loadings. For BA-estimation, the salient loadings were freely estimated without priors, and normally distributed priors for the non-salient loadings were estimated with a zero mean and a variance of σ²=0.01.

Table 1. Example for a population loading pattern for 3 factors, 5 salient loadings per factor of .80, and with cross loadings of .40

| variable | $\lambda_1$ | $\lambda_2$ | $\lambda_3$ |
|---|---|---|---|
| *x1* | **.80** | **.40** | .00 |
| *x2* | **.80** | .00 | .00 |
| *x3* | **.80** | .00 | .00 |
| *x4* | **.80** | .00 | .00 |
| *x5* | **.80** | .00 | .00 |
| *x6* | .00 | **.80** | **.40** |
| *x7* | .00 | **.80** | .00 |
| *x8* | .00 | **.80** | .00 |
| *x9* | .00 | **.80** | .00 |
| *x10* | .00 | **.80** | .00 |
| *x11* | **.40** | .00 | **.80** |
| *x12* | .00 | .00 | **.80** |
| *x13* | .00 | .00 | **.80** |
| *x14* | .00 | .00 | **.80** |
| *x15* | .00 | .00 | **.80** |

*Note.* Loadings > .00 are written in boldface.

The parameter-based determinacy coefficients were computed according to Equations 4, 6, and 8. Determinacy coefficients were averaged across all factors of each model. Bias was computed as the difference between the parameter-based determinacy coefficient (right side of Equations 4, 6, and 8) and the respective score-based determinacy coefficient (left side of Equations 4, 6, and 8). Determinacy coefficients according to Equation 4 and Equation 8 were computed for ML-, WLSMV- and BA-estimation. Parameters from ML- and WLSMV-estimation were entered into Equation 6. Overall, seven parameter-based determinacy coefficients and their bias were investigated.

*Results of the simulation for finite populations*

The model fit is reported in the Appendix (Table S1). The models comprising misfit had a moderate fit when estimated with ML- and WLSMV. The Comparative Fit Index (CFI) was the most sensitive index for indicating misfit for these models. The CFI indicates a moderate fit of the models based on the misspecification of cross-loadings, whereas the Root Mean Square Error of Approximation (RMSEA) and the Standardized Root Mean Square Residual (SRMR) indicate that the fit of all models comprising models with misfit is acceptable. With BA-estimation, the RMSEA and the CFI indicate perfect model fit, even for the models with misspecification. In these models, only the Posterior Predictive P-Value (PPP) indicates some misfit.

Regarding determinacy coefficients, the most obvious result is the substantial positive bias that occurs when WLSMV-parameters are inserted into Equations 4 and 8 (see Table 2). The bias of the remaining determinacy coefficients was small. Thus, the effects of model misfit and factor inter-correlations on bias could be neglected at the population level. In all but one conditions, the ML-based determinacy coefficients were similar to the WLSMV-based determinacy coefficients. This is notable because the observed variables had only two categories. Using the WLSMV-based factor score predictor does not improve the score-based determinacy coefficients. When there are model misfit and factor inter-correlations, the score-based determinacy coefficient for WLSMV-estimation was substantially smaller than the score-based determinacy coefficient for ML-estimation (see Table 2). This implies that from the perspective of determinacy coefficients, the use of WLSMV-based factor scores and determinacy coefficients is not necessarily optimal, even when the data are categorical. It is, moreover, interesting that the size of the determinacy coefficients for the correlation preserving factor score predictor was the same as the size of the determinacy coefficients for the best linear factor score predictor.

Bias of determinacy coefficients 10Table 2. Finite population determinacy coefficients and bias of parameter-based determinacy coefficients based on ML- and WLSMV/ML-estimation for variables with two categories

| | | | sl = .40 | | | | sl = .80 | | | |
| | | | $\phi = .00$ | | $\phi = .30$ | | $\phi = .00$ | | $\phi = .30$ | |
| estimation method | Equation | coefficient | nl = .00 | nl = .20 | nl = .00 | nl = .20 | nl = .00 | nl = .40 | nl = .00 | nl = .40 |
| --- | --- | --- | --- | --- | --- | --- | --- | --- | --- | --- |
| ML | 4 | $Cor(\hat{\xi}_{BL},\xi_u)$ | .60 | .60 | .62 | .63 | .86 | .86 | .86 | .86 |
| | | $\mathbf{P}_{BL}/\Delta\mathbf{P}_{BL}$ | .59 / -.01 | .59 / -.01 | .61 / -.01 | .65 / .02 | .86 / .00 | .86 / .00 | .87 / .01 | .89 / .03 |
| WLSMV | 4 | $Cor(\hat{\xi}_{BL},\xi_u)$ | .60 | .60 | .62 | .63 | .86 | .86 | .87 | .78 |
| | | $\mathbf{P}_{BL}/\Delta\mathbf{P}_{BL}$ | .68 / **.08** | .68 / **.08** | .70 / **.08** | .73 / **.10** | .92 / **.06** | .92 / **.06** | .92 / **.05** | .98 / **.20** |
| BA | 4 | $Cor(\hat{\xi}_{BL},\xi_u)$ | .60 | .61 | .62 | .64 | .86 | .86 | .86 | .87 |
| | | $\mathbf{P}_{BL}/\Delta\mathbf{P}_{BL}$ | .59 / -.01 | .60 / -.01 | .61 / -.01 | .64 / .00 | .86 / .00 | .87 / .01 | .87 / .01 | .89 / .02 |
| ML/WLSMV | 6 | $\mathbf{P}_{BLc}/\Delta\mathbf{P}_{BLc}$ | .59 / -.01 | .59 / -.01 | .61 / -.01 | .65 / .02 | .86 / .00 | .86 / .00 | .87 / .01 | .80 / .02 |
| ML | 8 | $Cor(\hat{\xi}_{CP},\xi_u)$ | .60 | .60 | .62 | .63 | .86 | .86 | .86 | .86 |
| | | $\mathbf{P}_{CP}/\Delta\mathbf{P}_{CP}$ | .59 / -.01 | .59 / -.01 | .61 / -.01 | .64 / .01 | .86 / .00 | .86 / .00 | .87 / .01 | .89 / .03 |
| WLSMV | 8 | $Cor(\hat{\xi}_{CP},\xi_u)$ | .60 | .60 | .61 | .63 | .86 | .86 | .86 | .87 |
| | | $\mathbf{P}_{CP}/\Delta\mathbf{P}_{CP}$ | .68 / **.08** | .68 / **.08** | .69 / **.08** | .64 / .01 | .92 / **.06** | .92 / **.06** | .92 / **.06** | .98 / **.11** |
| BA | 8 | $Cor(\hat{\xi}_{CP},\xi_u)$ | .60 | .60 | .62 | .64 | .86 | .86 | .86 | .87 |
| | | $\mathbf{P}_{CP}/\Delta\mathbf{P}_{CP}$ | .59 / -.01 | .60 / .00 | .61 / -.01 | .64 / .00 | .86 / .00 | .87 / .01 | .87 / .01 | .89 / .02 |

*Note.* Standard deviations are given in brackets; bias of parameter-based determinacy coefficients is marked with "Δ"; score-based determinacy coefficients and bias greater or equal .05 are written in boldface; ML = maximum-likelihood; WLSMV = weighted least squares, mean- and variance- adjusted $Cor(\hat{\xi}_{BL},\xi_u)$ = score-based determinacy coefficient based on the best-linear factor score predictor, which was directly computed in Mplus 8.4; $\mathbf{P}_{BL}$ = determinacy coefficient for the best linear predictor; $\mathbf{P}_{BLc}$ = determinacy coefficient for the categorical best linear predictor; $Cor(\hat{\xi}_{CP},\xi_u)$ = score-based determinacy coefficient based on the correlation-preserving factor score predictor; $\mathbf{P}_{CP}$ = determinacy coefficient for the correlation preserving predictor, sl = salient loading, $\phi$ = factor inter-correlation, nl = non-salient cross-loading; the finite population comprises 1,000,000 cases.

*Simulation for samples*

Except the determinacy coefficients based on WLSMV-parameters inserted into Equations 4 and 6, all other determinacy coefficients had minimal bias in the population. In the next step, the bias of these determinacy coefficients, except the coefficients solely based on WLSMV-parameters, was investigated in a simulation based on samples. In addition to varying sample size with $n \in \{300, 900\}$, two additional conditions that might enhance the effect of sampling error on determinacy coefficients were introduced: The number of factors $q \in \{3, 5\}$ and the number of observed variables with salient loadings per factor $p/q \in \{5, 10\}$. The remaining models and conditions were exactly the same as in the population-based simulation study. In total, this resulted in 2 ($q$) × 2 ($p/q$) × 2 ($sl$) × 2 ($cl$) × 2 ($\phi$) × 4 ($c$) × 2 ($n$) = 256 simulation conditions. For each condition, 1,000 samples were drawn from multivariate normal distributions, transformed into symmetric binomial distributions, and submitted to ML-estimation, WLSMV-estimation, and BA-estimation. When sampling error is investigated, it might be of interest to investigate the effect of the correction of determinacy coefficients for sampling error according to Budescu (1982). Therefore, the determinacy coefficients were investigated with and without Budescu's correction for sampling error. Determinacy coefficients and bias were averaged across all factors of each model and across all samples in each simulation condition.

*Results of the simulation for samples*

The fit of the models for the samples is presented in the Appendix (Table S5). As the population-based simulation showed that the WLSMV-based determinacy coefficients resulted in substantial overestimation, the focus of the sample-based simulation was on ML- and BA-estimation. Therefore, the fit was only reported for the latter estimation methods. As expected, the CFI and RMSEA indicated model fit for models without misfit in the population. The CFI indicated poor fit of the models based on misfit, small samples, and ML-estimation. However, the RMSEA indicated fit for these models as for the other models. For BA-estimation, the fit of all models was very good. Overall, the effect of the number of categories on model fit was very small.

As expected, the mean score-based and mean parameter-based determinacy coefficients based on ML- and WLSMV/ML-estimation were considerably larger when they were based on large salient loadings as compared to small salient loadings (see Table 3). The mean standard deviations were small, even for small salient loadings and small sample sizes. Overall, the parameter-based determinacy coefficients were slightly larger than the corresponding score-based determinacy coefficients, which corresponds to a small positive bias of the parameter-based determinacy coefficients. The positive bias was smaller when Budescu's (1982) correction for sampling error was performed. However, a substantial positive bias subsisted even after correction for sampling error for data based on two categories. The results for the determinacy coefficient based on parameters from WLSMV- and ML-estimation were similar to the results for the coefficient based solely on ML-estimation, even for the conditions based on variables with two categories, where the positive bias was most pronounced. The score-based determinacy coefficients for the correlation-preserving factor score predictor were similar to the score-based determinacy coefficients of the best-linear factor score predictor. The positive bias of the parameter-based determinacy coefficient for the correlation-preserving factor score predictor and the reduction of bias due to the correction for sampling error were similar to what was found for the best-linear factor score predictor. The largest positive bias of the mean parameter-based determinacy coefficients occurred for two categories and salient loadings of .80. In these conditions, the bias was greater than .05, even for coefficients that were corrected for sampling error (Table 3).

Table 3. Mean determinacy coefficients and bias of parameter-based determinacy coefficients based on ML- and WLSMV/ML-estimation for small and large salient loadings (*sl*), small and large samples (*n*), for variables with different numbers of categories (*c*)

| | | | ML-estimation | | | | WLSMV/ML-estimation | | | | ML-estimation | | | |
|---|---|---|---|---|---|---|---|---|---|---|---|---|---|---|
| *sl* | *n* | *c* | $Cor(\hat{\xi}_{BL},\xi_u)$ | $\mathbf{P}_{BL}$ | $/\Delta\mathbf{P}_{BL}$ | $\mathbf{P}^c_{BL}$ | $/\Delta\mathbf{P}^c_{BL}$ | $\mathbf{P}_{BLc}$ | $/\Delta\mathbf{P}_{BLc}$ | $\mathbf{P}^c_{BLc}$ | $/\Delta\mathbf{P}^c_{BLc}$ | $Cor(\hat{\xi}_{CP},\xi_u)$ | $\mathbf{P}_{CP}$ | $/\Delta\mathbf{P}_{CP}$ | $\mathbf{P}^c_{CP}$ | $/\Delta\mathbf{P}^c_{CP}$ |
| .40 | 300 | 2 | **.65 (.02)** | .71 (.02) /**.06** | .67 (.02) /.02 | .71 (.02) /**.06** | .67 (.02) /.02 | **.65 (.02)** | .71 (.02) /**.06** | .67 (.02) /.02 |
| | | 4 | **.72 (.02)** | .76 (.01) /.04 | .73 (.02) /.01 | .76 (.01) /.04 | .73 (.02) /.01 | **.72 (.02)** | .75 (.01) /.03 | .72 (.02) /.00 |
| | | 6 | **.74 (.02)** | .77 (.01) /.03 | .74 (.02) /.00 | .76 (.01) /.02 | .74 (.02) /.00 | **.74 (.02)** | .76 (.01) /.02 | .73 (.02) /-.01 |
| | | 8 | **.74 (.02)** | .77 (.01) /.03 | .74 (.01) /.00 | .77 (.01) /.03 | .74 (.02) /.00 | **.74 (.02)** | .77 (.01) /.03 | .74 (.01) /.00 |
| | 900 | 2 | **.67 (.01)** | .69 (.01) /.02 | .68 (.01) /.01 | .69 (.01) /.02 | .68 (.01) /.01 | **.67 (.01)** | .69 (.01) /.02 | .68 (.01) /.01 |
| | | 4 | **.73 (.02)** | .75 (.01) /.02 | .74 (.01) /.01 | .75 (.01) /.02 | .74 (.01) /.01 | **.73 (.02)** | .75 (.01) /.02 | .74 (.01) /.01 |
| | | 6 | **.75 (.01)** | .76 (.01) /.01 | .75 (.01) /.00 | .76 (.01) /.01 | .75 (.01) /.00 | **.74 (.01)** | .76 (.01) /.02 | .75 (.01) /.01 |
| | | 8 | **.75 (.01)** | .76 (.01) /.01 | .75 (.01) /.00 | .76 (.01) /.01 | .75 (.01) /.00 | **.75 (.01)** | .76 (.01) /.01 | .75 (.01) /.00 |
| .80 | 300 | 2 | **.85 (.06)** | .93 (.00) /**.08** | .92 (.00) /**.07** | .93 (.00) /**.08** | .92 (.00) /**.07** | **.85 (.06)** | .93 (.00) /**.08** | .92 (.00) /**.07** |
| | | 4 | **.94 (.00)** | .95 (.00) /.01 | .95 (.00) /.01 | .95 (.00) /.01 | .95 (.00) /.01 | **.94 (.00)** | .95 (.00) /.01 | .95 (.00) /.01 |
| | | 6 | **.95 (.00)** | .96 (.00) /.01 | .96 (.00) /.01 | .96 (.00) /.01 | .96 (.00) /.01 | **.95 (.00)** | .96 (.00) /.01 | .96 (.00) /.01 |
| | | 8 | **.95 (.00)** | .96 (.00) /.01 | .96 (.00) /.01 | .96 (.00) /.01 | .96 (.00) /.01 | **.95 (.00)** | .96 (.00) /.01 | .96 (.00) /.01 |
| | 900 | 2 | **.87 (.04)** | .93 (.00) /**.06** | .92 (.00) /**.05** | .93 (.00) /**.06** | .92 (.00) /**.05** | **.87 (.04)** | .93 (.00) /**.06** | .92 (.00) /**.05** |
| | | 4 | **.94 (.00)** | .95 (.00) /.01 | .95 (.00) /.01 | .95 (.00) /.01 | .95 (.00) /.01 | **.94 (.00)** | .95 (.00) /.01 | .95 (.00) /.01 |
| | | 6 | **.95 (.00)** | .96 (.00) /.01 | .96 (.00) /.01 | .96 (.00) /.01 | .96 (.00) /.01 | **.95 (.00)** | .96 (.00) /.01 | .96 (.00) /.01 |
| | | 8 | **.95 (.00)** | .96 (.00) /.01 | .96 (.00) /.01 | .96 (.00) /.01 | .96 (.00) /.01 | **.95 (.00)** | .96 (.00) /.01 | .96 (.00) /.01 |

*Note.* Standard deviations are given in brackets; bias of parameter-based determinacy coefficients is marked with "Δ"; score-based determinacy coefficients and bias greater or equal .05 are written in boldface; ML = maximum-likelihood; WLSMV = weighted least squares, mean- and variance- adjusted $Cor(\hat{\xi}_{BL},\xi_u)$ = score-based determinacy coefficient based on the best-linear factor score predictor; $\mathbf{P}_{BL}$ = determinacy coefficient for the best linear predictor; $\mathbf{P}^c_{BL} = \mathbf{P}_{BL}$, corrected for sampling error; $\mathbf{P}_{BLc}$ = determinacy coefficient for the categorical best linear predictor; $\mathbf{P}^c_{BLc} = \mathbf{P}_{BLc}$, corrected for sampling error; $Cor(\hat{\xi}_{CP},\xi_u)$ = score-based determinacy coefficient based on the correlation-preserving factor score predictor; $\mathbf{P}_{CP}$ = determinacy coefficient for the correlation preserving predictor; $\mathbf{P}^c_{CP} = \mathbf{P}_{CP}$, corrected for sampling error.

The results for BA-estimation were similar to the results based on ML- and WLSMV/ML-estimation (see Table 4). Overall, the parameter-based determinacy coefficients were slightly larger than the corresponding score-based determinacy coefficients. The positive bias of the parameter-based determinacy coefficients was most pronounced for conditions based on variables with two categories, and the correction for sampling error reduced the positive bias of the parameter-based determinacy coefficients. As for ML- and WLSMV/ML-estimation, the largest positive bias of the mean parameter-based determinacy coefficients occurred for two categories and salient loadings of .80. In these conditions, the bias was again greater than .05, even for coefficients corrected for sampling error (Table 4).

From the remaining conditions, the condition with non-zero cross-loadings and non-zero factor inter-correlations induced the most substantial bias. Therefore, the mean determinacies for this condition were presented in the Appendix. A bias greater than .10 occurred for two categories, salient loadings of .80 for the determinacy coefficients based on ML- and WLSMV/ML-parameter estimates (see Table S3). Similar results were found for BA-estimation. A bias of at least .10 occurred for two categories, salient loadings of .80 for the determinacy coefficients based on Bayes parameter estimates (see Table S4). In contrast, the bias of determinacy coefficients based on parameter estimates based on ML-, WLSMV/ML-, and BA-estimation in the conditions with zero non-salient loadings and zero factor inter-correlations was smaller than .05, even in the conditions based on two categories and salient loadings of .80 (see Appendix, Tables S5 and S6).

Table 4. Mean determinacy coefficients and bias of parameter-based determinacy coefficients based on Bayes estimation for small and large salient loadings (*sl*), small and large samples (*n*), for variables with different number of categories (*c*)

| sl | n | c | $Cor(\hat{\xi}_{BL},\xi_u)$ | $\mathbf{P}_{BL}$ | $/\Delta\mathbf{P}_{BL}$ | $\mathbf{P}^c_{BL}$ | $/\Delta\mathbf{P}^c_{BL}$ | $Cor(\hat{\xi}_{CP},\xi_u)$ | $\mathbf{P}_{CP}$ | $/\Delta\mathbf{P}_{CP}$ | $\mathbf{P}^c_{CP}$ | $/\Delta\mathbf{P}^c_{CP}$ |
|---|---|---|---|---|---|---|---|---|---|---|---|---|
| .40 | 300 | 2 | **.63 (.02)** | .67 (.03) | /.04 | .63 (.04) | /.00 | **.62 (.02)** | .67 (.03) | /.05 | .62 (.04) | /.00 |
|  |  | 4 | **.72 (.02)** | .75 (.02) | /.03 | .73 (.02) | /.01 | **.72 (.02)** | .75 (.02) | /.03 | .72 (.02) | /.00 |
|  |  | 6 | **.74 (.02)** | .76 (.01) | /.02 | .74 (.02) | /.00 | **.73 (.02)** | .76 (.01) | /.03 | .73 (.02) | /.00 |
|  |  | 8 | **.74 (.02)** | .77 (.01) | /.03 | .74 (.02) | /.00 | **.74 (.02)** | .76 (.01) | /.02 | .74 (.02) | /.00 |
|  | 900 | 2 | **.66 (.01)** | .68 (.01) | /.02 | .66 (.02) | /.00 | **.65 (.01)** | .67 (.02) | /.02 | .66 (.02) | /.01 |
|  |  | 4 | **.73 (.01)** | .75 (.01) | /.02 | .74 (.01) | /.01 | **.73 (.01)** | .75 (.01) | /.02 | .74 (.01) | /.01 |
|  |  | 6 | **.75 (.01)** | .76 (.01) | /.01 | .75 (.01) | /.00 | **.74 (.01)** | .76 (.01) | /.02 | .75 (.01) | /.01 |
|  |  | 8 | **.75 (.01)** | .76 (.01) | /.01 | .76 (.01) | /.01 | **.75 (.01)** | .76 (.01) | /.01 | .75 (.01) | /.00 |
| .80 | 300 | 2 | **.87 (.01)** | .93 (.00) | **/.06** | .93 (.00) | **/.06** | **.87 (.01)** | .93 (.00) | **/.06** | .93 (.00) | **/.06** |
|  |  | 4 | **.95 (.00)** | .96 (.00) | /.01 | .95 (.00) | /.00 | **.95 (.00)** | .96 (.00) | /.01 | .95 (.00) | /.00 |
|  |  | 6 | **.96 (.00)** | .96 (.00) | /.00 | .96 (.00) | /.00 | **.96 (.00)** | .96 (.00) | /.00 | .96 (.00) | /.00 |
|  |  | 8 | **.96 (.00)** | .96 (.00) | /.00 | .96 (.00) | /.00 | **.96 (.00)** | .96 (.00) | /.00 | .96 (.00) | /.00 |
|  | 900 | 2 | **.87 (.00)** | .93 (.00) | **/.06** | .93 (.00) | **/.06** | **.87 (.00)** | .93 (.00) | **/.06** | .93 (.00) | **/.06** |
|  |  | 4 | **.95 (.00)** | .96 (.00) | /.01 | .95 (.00) | /.00 | **.95 (.00)** | .96 (.00) | /.01 | .95 (.00) | /.00 |
|  |  | 6 | **.96 (.00)** | .96 (.00) | /.00 | .96 (.00) | /.00 | **.96 (.00)** | .96 (.00) | /.00 | .96 (.00) | /.00 |
|  |  | 8 | **.96 (.00)** | .96 (.00) | /.00 | .96 (.00) | /.00 | **.96 (.00)** | .96 (.00) | /.00 | .96 (.00) | /.00 |

*Note.* Standard deviations are given in brackets; bias of parameter-based determinacy coefficients is marked with "Δ"; score-based determinacy coefficients and bias greater or equal .05 are given in bold face; $Cor(\hat{\xi}_{BL},\xi_u)$ = score-based determinacy coefficient based on the best-linear factor score predictor; $\mathbf{P}_{BL}$ = determinacy coefficient for the best linear predictor; $\mathbf{P}^c_{BL} = \mathbf{P}_{BL}$, corrected for sampling error; $Cor(\hat{\xi}_{CP},\xi_u)$ = score-based determinacy coefficient based on the correlation-preserving factor score predictor; $\mathbf{P}_{CP}$ = determinacy coefficient for the correlation preserving predictor; $\mathbf{P}^c_{CP} = \mathbf{P}_{CP}$, corrected for sampling error.

**Discussion**

The present study investigated the size and bias of determinacy coefficients for the best linear factor score predictor and for a correlation-preserving factor score predictor for WLSMV-, ML-, and BA-estimation under different conditions. The advantage of the present approach based on a simulation study is that the factor score predictors could be correlated with the factor scores used for generating the observed variables. The resulting score-based determinacy coefficients are a bench-mark for the parameter-based determinacy coefficients. Therefore, bias of parameter-based determinacy coefficients was computed as the difference between parameter-based determinacy coefficients and the corresponding score-based determinacy coefficients. This is relevant because in empirical settings, the score-based determinacy coefficients are unavailable.

First, a simulation based on finite populations was performed for three-factor models based on observed variables with two categories. When the WLSMV-parameters were entered into the Equations for determinacy coefficients, the parameter-based determinacy coefficients overestimated the size of the score-based determinacy coefficients. This indicates that WLSMV-parameter-based coefficients are estimates for the determinacy that would have occurred if the factor score predictors and the factor scores were both continuous (Beauducel & Hilger, 2017b). The determinacy coefficient based on combined ML/WLSMV-estimates (Beauducel & Hilger, 2017b) was a useful alternative to the WLSMV-based coefficient for CFA based on categorical variables and more than one factor.

The determinacy coefficients computed from ML-parameters and the determinacy coefficients computed from BA-parameters had minimal bias in the population, even for conditions with correlated factors and model misfit. The determinacy coefficients for the best linear factor score predictor were similar to the determinacy coefficients for the correlation-preserving factor score predictor. Moreover, the determinacy coefficient computed from combined ML- and WLSMV-parameters according to Equation 6 had a rather small bias when compared with the score-based determinacy coefficients computed from the categorical scores with Mplus 8.4. Therefore, this determinacy coefficient might be of interest when categorical factor score predictors are computed.

The sample-based simulation study revealed that parameter-based determinacy coefficients based on the estimation methods investigated here resulted in moderate but consistent positive bias when based on two categories and large salient loadings. The positive bias was small but subsisted when Budescu's (1982) correction for sampling error was performed. However, in light of the results of the simulation study for samples, Budescu's correction is recommended. The bias was substantial when two-categories and large salient loadings co-occurred with non-zero non-salient loadings (misfit) and non-zero factor inter-correlations. For these conditions, bias was large even in large samples. As model fit did not decrease substantially with the number of categories, substantial bias of parameter-based determinacy coefficients is likely to occur in conditions based on two categories that might be regarded as acceptable for CFA. In contrast, only small positive bias of determinacy coefficients occurred with four and more categories of observed variables with ML-estimation, and bias can even be neglected with four and more categories with BA-estimation. These results were found for the determinacy coefficients of the best linear factor score predictor as well as for the determinacy coefficients of the correlation-preserving factor score predictor. The similarity of determinacy coefficients and bias for the best linear factor score predictor and for the correlation-preserving factor score predictor corroborates recent results on this issue (Beauducel, Hilger & Kuhl, in press).

As a limitation of the present simulation study, it should be noted that only constant salient population loadings, moderate factor inter-correlations, and symmetric categorical variables were investigated. Simulation studies based on more complex loading patterns, higher factor inter-correlations, and asymmetric categorical variables may lead to further insights. Moreover, further research might investigate determinacy coefficients for conditionally unbiased factor score predictors (Krijnen, Wansbeek, & Ten Berge, 1996) and for other estimation methods of model parameters.

Nevertheless, the present results support the following recommendations when factor score predictors are estimated: Budescu's (1982) correction for sampling error of determinacy coefficients should be performed, ML- or BA-parameters should be used, and data should be based on at least four categories to avoid overestimation of parameter-based determinacy coefficients. Overestimation of determinacy coefficients was less pronounced for BA-estimation than for ML-estimation. Finally, bias and size of determinacy coefficients for the best linear factor score predictor and the correlation-preserving factor score predictor were similar. Thus, determinacy coefficients do not preclude that the best linear factor score predictor should be preferred over correlation-preserving factor score predictor.

**Acknowledgement**

This study was funded by the German Research Foundation (DFG), BE 2443/18-1.

# Appendix

Table S1. Fit of population models.

| est. method | fit index | sl = .40 ϕ = .00 nl = .00 | sl = .40 ϕ = .00 nl = .20 | sl = .40 ϕ = .30 nl = .00 | sl = .40 ϕ = .30 nl = .20 | sl = .80 ϕ = .00 nl = .00 | sl = .80 ϕ = .00 nl = .40 | sl = .80 ϕ = .30 nl = .00 | sl = .80 ϕ = .30 nl = .40 |
|---|---|---|---|---|---|---|---|---|---|
| ML | $\chi^2(df), p$ | 77.57(90), .82 | 29081.96(90),<.01 | 75.29(87), .81 | 17571.40(87), .00 | 83.94(90), .66 | 374146.77(90),<.01 | 71.86(87), .88 | 447535.65(87),<.01 |
|  | RMSEA | .000 | .018 | .000 | .014 | .000 | .064 | .000 | .072 |
|  | CFI | 1.000 | .901 | 1.000 | .955 | 1.000 | .913 | 1.000 | .918 |
|  | SRMR | .001 | .017 | .001 | .012 | .001 | .069 | .001 | .058 |
| WLSMV | $\chi^2(df), p$ | 77.15(90), .83 | 39375.62(90),<.01 | 76.05(87), .79 | 20928.14(87),<.01 | 81.41(90), .73 | 431695.79(90),<.01 | 78.05(87), .74 | 622554.51(87),<.01 |
|  | RMSEA | .000 | .021 | .000 | .015 | .000 | .069 | .000 | .085 |
|  | CFI | 1.000 | .887 | 1.000 | .960 | 1.000 | .934 | 1.000 | .935 |
|  | SRMR | .001 | .027 | .001 | .018 | .001 | .107 | .001 | .079 |
| BA | PPP | .67 | .38 | .72 | .29 | .75 | .43 | .77 | <.01 |
|  | RMSEA | .000 | .000 | .000 | .000 | .000 | .000 | .000 | .003 |
|  | CFI | 1.000 | 1.000 | 1.000 | 1.000 | 1.000 | 1.000 | 1.000 | 1.000 |

*Note.* RMSEA = Root Mean Square Error of Approximation, CFI = Comparative Fit Index; SRMR = Standardized Root Mean Square Residual, PPP = Posterior Predictive P-Value, *sl* = salient loading, ϕ = factor inter-correlation, *nl* = non-salient cross-loading.

Table S2. Fit of sample models.

| nl (misfit) | n | c | ML | | BA | |
|---|---|---|---|---|---|---|
| | | | CFI | RMSEA | CFI | RMSEA |
| No | 300 | 2 | .948 (.044) | .015 (.008) | .965 (.032) | .014 (.008) |
| | | 4 | .970 (.028) | .012 (.009) | .984 (.019) | .009 (.008) |
| | | 6 | .973 (.025) | .011 (.009) | .987 (.017) | .008 (.008) |
| | | 8 | .974 (.024) | .012 (.009) | .988 (.016) | .008 (.008) |
| | 900 | 2 | .987 (.015) | .007 (.005) | .991 (.012) | .007 (.005) |
| | | 4 | .993 (.009) | .005 (.005) | .996 (.006) | .004 (.005) |
| | | 6 | .994 (.008) | .005 (.005) | .997 (.005) | .004 (.005) |
| | | 8 | .994 (.007) | .005 (.005) | .998 (.005) | .003 (.004) |
| Yes | 300 | 2 | .887 (.048) | .037 (.007) | .961 (.029) | .019 (.008) |
| | | 4 | .894 (.034) | .049 (.006) | .982 (.018) | .013 (.008) |
| | | 6 | .892 (.031) | .054 (.006) | .985 (.017) | .012 (.008) |
| | | 8 | .890 (.030) | .056 (.006) | .984 (.017) | .012 (.009) |
| | 900 | 2 | .917 (.023) | .035 (.003) | .983 (.011) | .013 (.005) |
| | | 4 | .911 (.015) | .047 (.003) | .994 (.006) | .008 (.005) |
| | | 6 | .906 (.014) | .052 (.003) | .996 (.005) | .006 (.005) |
| | | 8 | .904 (.014) | .054 (.003) | .997 (.005) | .006 (.005) |

Table S3. Mean determinacy coefficients and bias of parameter-based determinacy coefficients based on ML- and WLSMV/ML-estimation for small and large salient loadings (*sl*), small and large samples (*n*), for variables with different numbers of categories (*c*) for models with non-zero cross-loadings (misfit) and non-zero factor inter-correlations

| | | | ML-estimation | | | | | WLSMV/ML-estimation | | | | ML-estimation | | | |
|---|---|---|---|---|---|---|---|---|---|---|---|---|---|---|---|
| *sl* | *n* | *c* | $Cor(\hat{\xi}_{BL},\xi_u)$ | $\mathbf{P}_{BL}$ | $/\Delta\mathbf{P}_{BL}$ | $\mathbf{P}^c_{BL}$ | $/\Delta\mathbf{P}^c_{BL}$ | $\mathbf{P}_{BLc}$ | $/\Delta\mathbf{P}_{BLc}$ | $\mathbf{P}^c_{BLc}$ | $/\Delta\mathbf{P}^c_{BLc}$ | $Cor(\hat{\xi}_{CP},\xi_u)$ | $\mathbf{P}_{CP}$ | $/\Delta\mathbf{P}_{CP}$ | $\mathbf{P}^c_{CP}$ | $/\Delta\mathbf{P}^c_{CP}$ |
| | | 2 | **.67 (.02)** | .74 (.02) | /**.07** | .71 (.02) | /.04 | .74 (.02) | /**.06** | .70 (.02) | /.03 | **.67 (.02)** | .73 (.02) | /**.06** | .70 (.02) | /.02 |
| | 300 | 4 | **.74 (.02)** | .78 (.01) | /.05 | .76 (.02) | /.02 | .78 (.01) | /.04 | .76 (.02) | /.02 | **.74 (.02)** | .78 (.01) | /.04 | .75 (.02) | /.01 |
| | | 6 | **.75 (.02)** | .79 (.01) | /.04 | .77 (.01) | /.02 | .79 (.01) | /.04 | .77 (.02) | /.02 | **.75 (.02)** | .79 (.01) | /.03 | .76 (.01) | /.01 |
| .40 | | 8 | **.75 (.02)** | .80 (.01) | /.04 | .77 (.01) | /.02 | .79 (.01) | /.04 | .77 (.01) | /.02 | **.76 (.02)** | .79 (.01) | /.03 | .76 (.01) | /.01 |
| | | 2 | **.69 (.01)** | .73 (.01) | /.04 | .72 (.01) | /.03 | .72 (.01) | /.03 | .71 (.01) | /.02 | **.69 (.01)** | .72 (.01) | /.03 | .71 (.01) | /.02 |
| | 900 | 4 | **.75 (.01)** | .78 (.01) | /.03 | .77 (.01) | /.02 | .77 (.01) | /.03 | .77 (.01) | /.02 | **.75 (.01)** | .77 (.01) | /.02 | .76 (.01) | /.01 |
| | | 6 | **.76 (.01)** | .79 (.01) | /.03 | .78 (.01) | /.02 | .78 (.01) | /.03 | .78 (.01) | /.02 | **.76 (.01)** | .78 (.01) | /.02 | .77 (.01) | /.01 |
| | | 8 | **.76 (.01)** | .79 (.01) | /.03 | .78 (.01) | /.02 | .79 (.01) | /.03 | .78 (.01) | /.02 | **.76 (.01)** | .78 (.01) | /.02 | .77 (.01) | /.01 |
| | | 2 | **.77 (.20)** | .95 (.00) | /**.18** | .95 (.00) | /**.18** | .95 (.00) | /**.18** | .95 (.00) | /**.18** | **.77 (.20)** | .95 (.00) | /**.18** | .95 (.00) | /**.17** |
| | 300 | 4 | **.92 (.01)** | .97 (.00) | /.05 | .97 (.00) | /.05 | .97 (.00) | /.05 | .97 (.00) | /.05 | **.92 (.01)** | .97 (.00) | /.05 | .97 (.00) | /.04 |
| | | 6 | **.93 (.00)** | .98 (.00) | /.04 | .97 (.00) | /.04 | .98 (.00) | /.04 | .97 (.00) | /.04 | **.93 (.00)** | .98 (.00) | /.04 | .97 (.00) | /.04 |
| .80 | | 8 | **.93 (.00)** | .98 (.00) | /.04 | .98 (.00) | /.04 | .98 (.00) | /.04 | .98 (.00) | /.04 | **.94 (.00)** | .98 (.00) | /.04 | .98 (.00) | /.04 |
| | | 2 | **.82 (.11)** | .95 (.00) | /**.13** | .95 (.00) | /**.13** | .95 (.00) | /**.13** | .95 (.00) | /**.13** | **.82 (.11)** | .95 (.00) | /**.13** | .95 (.00) | /**.13** |
| | 900 | 4 | **.92 (.00)** | .97 (.00) | /.05 | .97 (.00) | /.05 | .97 (.00) | /.05 | .97 (.00) | /.05 | **.93 (.00)** | .97 (.00) | /.04 | .97 (.00) | /.04 |
| | | 6 | **.93 (.00)** | .98 (.00) | /.04 | .98 (.00) | /.04 | .98 (.00) | /.04 | .98 (.00) | /.04 | **.94 (.00)** | .98 (.00) | /.04 | .98 (.00) | /.04 |
| | | 8 | **.94 (.00)** | .98 (.00) | /.04 | .98 (.00) | /.04 | .98 (.00) | /.04 | .98 (.00) | /.04 | **.94 (.00)** | .98 (.00) | /.04 | .98 (.00) | /.04 |

*Note.* Standard deviations are given in brackets; bias of parameter-based determinacy coefficients is marked with "Δ"; score-based determinacy coefficients and bias greater or equal .05 are given in bold face; ML = maximum-likelihood; WLSMV = weighted least squares, mean- and variance- adjusted ; $Cor(\hat{\xi}_{BL},\xi_u)$ = score-based determinacy coefficient based on the best-linear factor score predictor; $\mathbf{P}_{BL}$ = determinacy coefficient for the best linear predictor; $\mathbf{P}^c_{BL}$ = $\mathbf{P}_{BL}$, corrected for sampling error; $\mathbf{P}_{BLc}$ = determinacy coefficient for the categorical best linear predictor; $\mathbf{P}^c_{BLc}$ = $\mathbf{P}_{BLc}$, corrected for sampling error; $Cor(\hat{\xi}_{CP},\xi_u)$ = score-based determinacy coefficient based on the correlation-preserving factor score predictor; $\mathbf{P}_{CP}$ = determinacy coefficient for the correlation preserving predictor; $\mathbf{P}^c_{CP}$ = $\mathbf{P}_{CP}$, corrected for sampling error.

Table S4. Mean determinacy coefficients and bias of parameter-based determinacy coefficients based on Bayes estimation for small and large salient loadings (*sl*), small and large samples (*n*), for variables with different numbers of categories (*c*) for models with non-zero cross-loadings (misfit) and non-zero factor inter-correlations

| *sl* | *n* | *c* | $Cor(\hat{\xi}_{BL},\xi_u)$ | $\mathbf{P}_{BL}$ | $/\Delta\mathbf{P}_{BL}$ | $\mathbf{P}^c_{BL}$ | $/\Delta\mathbf{P}^c_{BL}$ | $Cor(\hat{\xi}_{CP},\xi_u)$ | $\mathbf{P}_{CP}$ | $/\Delta\mathbf{P}_{CP}$ | $\mathbf{P}^c_{CP}$ | $/\Delta\mathbf{P}^c_{CP}$ |
|---|---|---|---|---|---|---|---|---|---|---|---|---|
| .40 | 300 | 2 | **.66 (.02)** | .70 (.03) | /.04 | .66 (.04) | /.00 | **.64 (.02)** | .69 (.03) | **/.05** | .65 (.04) | /.01 |
|  |  | 4 | **.74 (.02)** | .78 (.01) | /.04 | .75 (.02) | /.01 | **.74 (.02)** | .77 (.01) | /.04 | .75 (.02) | /.01 |
|  |  | 6 | **.75 (.02)** | .79 (.01) | /.04 | .76 (.02) | /.01 | **.75 (.02)** | .78 (.01) | /.03 | .76 (.02) | /.01 |
|  |  | 8 | **.76 (.02)** | .79 (.01) | /.03 | .77 (.01) | /.01 | **.76 (.01)** | .79 (.01) | /.03 | .76 (.01) | /.00 |
|  | 900 | 2 | **.68 (.01)** | .71 (.02) | /.02 | .69 (.02) | /.01 | **.66 (.01)** | .70 (.02) | /.03 | .69 (.02) | /.02 |
|  |  | 4 | **.75 (.01)** | .78 (.01) | /.03 | .77 (.01) | /.02 | **.75 (.01)** | .77 (.01) | /.02 | .76 (.01) | /.02 |
|  |  | 6 | **.76 (.01)** | .79 (.01) | /.02 | .78 (.01) | /.02 | **.76 (.01)** | .78 (.01) | /.02 | .77 (.01) | /.01 |
|  |  | 8 | **.77 (.01)** | .79 (.01) | /.02 | .78 (.01) | /.01 | **.77 (.01)** | .78 (.01) | /.02 | .78 (.01) | /.01 |
| .80 | 300 | 2 | **.85 (.01)** | .96 (.00) | **/.11** | .95 (.00) | **/.10** | **.85 (.01)** | .96 (.00) | **/.11** | .95 (.00) | **/.10** |
|  |  | 4 | **.95 (.00)** | .97 (.00) | /.03 | .97 (.00) | /.02 | **.95 (.00)** | .97 (.00) | /.02 | .97 (.00) | /.02 |
|  |  | 6 | **.96 (.00)** | .98 (.00) | /.02 | .98 (.00) | /.01 | **.97 (.00)** | .98 (.00) | /.01 | .98 (.00) | /.01 |
|  |  | 8 | **.97 (.00)** | .98 (.00) | /.01 | .98 (.00) | /.01 | **.97 (.00)** | .98 (.00) | /.01 | .98 (.00) | /.01 |
|  | 900 | 2 | **.85 (.01)** | .95 (.00) | **/.10** | .95 (.00) | **/.10** | **.85 (.01)** | .95 (.00) | **/.10** | .95 (.00) | **/.10** |
|  |  | 4 | **.95 (.00)** | .97 (.00) | /.02 | .97 (.00) | /.02 | **.95 (.00)** | .97 (.00) | /.02 | .97 (.00) | /.02 |
|  |  | 6 | **.97 (.00)** | .98 (.00) | /.01 | .98 (.00) | /.01 | **.97 (.00)** | .98 (.00) | /.01 | .98 (.00) | /.01 |
|  |  | 8 | **.97 (.00)** | .98 (.00) | /.01 | .98 (.00) | /.01 | **.97 (.00)** | .98 (.00) | /.01 | .98 (.00) | /.01 |

*Note.* Standard deviations are given in brackets; bias of parameter-based determinacy coefficients is marked with "Δ"; score-based determinacy coefficients and bias greater or equal .05 are given in bold face; $Cor(\hat{\xi}_{BL},\xi_u)$ = score-based determinacy coefficient based on the best-linear factor score predictor; $\mathbf{P}_{BL}$ = determinacy coefficient for the best linear predictor; $\mathbf{P}^c_{BL} = \mathbf{P}_{BL}$, corrected for sampling error; $Cor(\hat{\xi}_{CP},\xi_u)$ = score-based determinacy coefficient based on the correlation-preserving factor score predictor; $\mathbf{P}_{CP}$ = determinacy coefficient for the correlation preserving predictor; $\mathbf{P}^c_{CP} = \mathbf{P}_{CP}$, corrected for sampling error.

Table S5. Mean determinacy coefficients and bias of parameter-based determinacy coefficients based on ML- and WLSMV/ML-estimation for small and large salient loadings (*sl*), small and large samples (*n*), for variables with different number of categories (*c*) for models with zero cross-loadings and zero factor inter-correlations

| | | | ML-estimation | | | | | WLSMV/ML-estimation | | | | ML-estimation | | | | |
|---|---|---|---|---|---|---|---|---|---|---|---|---|---|---|---|---|
| *sl* | *n* | *c* | $Cor(\hat{\xi}_{BL},\xi_u)$ | $\mathbf{P}_{BL}$ | $/\Delta\mathbf{P}_{BL}$ | $\mathbf{P}^c_{BL}$ | $/\Delta\mathbf{P}^c_{BL}$ | $\mathbf{P}_{BLc}$ | $/\Delta\mathbf{P}_{BLc}$ | $\mathbf{P}^c_{BLc}$ | $/\Delta\mathbf{P}^c_{BLc}$ | $Cor(\hat{\xi}_{CP},\xi_u)$ | $\mathbf{P}_{CP}$ | $/\Delta\mathbf{P}_{CP}$ | $\mathbf{P}^c_{CP}$ | $/\Delta\mathbf{P}^c_{CP}$ |
| .40 | 300 | 2 | **.64 (.02)** | .69 (.02) | /**.05** | .65 (.02) | /.01 | .69 (.02) | /**.05** | .65 (.02) | /.01 | **.64 (.02)** | .69 (.02) | /**.05** | .65 (.02) | /.01 |
| | | 4 | **.71 (.03)** | .74 (.01) | /.03 | .71 (.02) | /-.01 | .74 (.01) | /.03 | .71 (.02) | /-.01 | **.71 (.03)** | .74 (.01) | /.03 | .71 (.02) | /-.01 |
| | | 6 | **.73 (.01)** | .75 (.01) | /.02 | .72 (.02) | /-.01 | .75 (.01) | /.02 | .72 (.02) | /-.01 | **.73 (.01)** | .75 (.01) | /.02 | .72 (.02) | /-.01 |
| | | 8 | **.73 (.01)** | .75 (.01) | /.02 | .73 (.02) | /-.01 | .75 (.01) | /.02 | .73 (.02) | /-.01 | **.73 (.01)** | .75 (.01) | /.02 | .73 (.02) | /-.01 |
| | 900 | 2 | **.66 (.01)** | .67 (.01) | /.02 | .66 (.01) | /.00 | .67 (.01) | /.02 | .66 (.01) | /.00 | **.66 (.01)** | .67 (.01) | /.02 | .66 (.01) | /.00 |
| | | 4 | **.72 (.02)** | .73 (.01) | /.01 | .72 (.01) | /.00 | .73 (.01) | /.01 | .72 (.01) | /.00 | **.72 (.02)** | .73 (.01) | /.01 | .72 (.01) | /.00 |
| | | 6 | **.74 (.01)** | .74 (.01) | /.01 | .73 (.01) | /.00 | .74 (.01) | /.01 | .73 (.01) | /.00 | **.74 (.01)** | .74 (.01) | /.01 | .73 (.01) | /.00 |
| | | 8 | **.74 (.01)** | .75 (.01) | /.01 | .74 (.01) | /.00 | .75 (.01) | /.01 | .74 (.01) | /.00 | **.74 (.01)** | .75 (.01) | /.01 | .74 (.01) | /.00 |
| .80 | 300 | 2 | **.88 (.01)** | .92 (.00) | /.04 | .91 (.00) | /.03 | .92 (.00) | /.04 | .91 (.00) | /.03 | **.88 (.01)** | .92 (.00) | /.04 | .91 (.00) | /.03 |
| | | 4 | **.94 (.00)** | .95 (.00) | /.00 | .94 (.00) | /.00 | .95 (.00) | /.00 | .94 (.00) | /.00 | **.94 (.00)** | .95 (.00) | /.00 | .94 (.00) | /.00 |
| | | 6 | **.95 (.00)** | .95 (.00) | /.00 | .95 (.00) | /.00 | .95 (.00) | /.00 | .95 (.00) | /.00 | **.95 (.00)** | .95 (.00) | /.00 | .95 (.00) | /.00 |
| | | 8 | **.95 (.00)** | .96 (.00) | /.00 | .95 (.00) | /.00 | .96 (.00) | /.00 | .95 (.00) | /.00 | **.95 (.00)** | .96 (.00) | /.00 | .95 (.00) | /.00 |
| | 900 | 2 | **.88 (.02)** | .92 (.00) | /.04 | .92 (.00) | /.04 | .92 (.00) | /.04 | .92 (.00) | /.04 | **.88 (.02)** | .92 (.00) | /.04 | .92 (.00) | /.04 |
| | | 4 | **.94 (.00)** | .95 (.00) | /.00 | .95 (.00) | /.00 | .95 (.00) | /.00 | .95 (.00) | /.00 | **.94 (.00)** | .95 (.00) | /.00 | .95 (.00) | /.00 |
| | | 6 | **.95 (.00)** | .95 (.00) | /.00 | .95 (.00) | /.00 | .95 (.00) | /.00 | .95 (.00) | /.00 | **.95 (.00)** | .95 (.00) | /.00 | .95 (.00) | /.00 |
| | | 8 | **.95 (.00)** | .96 (.00) | /.00 | .95 (.00) | /.00 | .96 (.00) | /.00 | .95 (.00) | /.00 | **.95 (.00)** | .96 (.00) | /.00 | .95 (.00) | /.00 |

*Note.* Standard deviations are given in brackets; bias of parameter-based determinacy coefficients is marked with "Δ"; score-based determinacy coefficients and bias greater or equal .05 are given in bold face; ML = maximum-likelihood; WLSMV = weighted least squares, mean- and variance- adjusted; $Cor(\hat{\xi}_{BL},\xi_u)$ = score-based determinacy coefficient based on the best-linear factor score predictor; $\mathbf{P}_{BL}$ = determinacy coefficient for the best linear predictor; $\mathbf{P}^c_{BL} = \mathbf{P}_{BL}$, corrected for sampling error; $\mathbf{P}_{BLc}$ = determinacy coefficient for the categorical best linear predictor; $\mathbf{P}^c_{BLc} = \mathbf{P}_{BLc}$, corrected for sampling error; $Cor(\hat{\xi}_{CP},\xi_u)$ = score-based determinacy coefficient based on the correlation-preserving factor score predictor; $\mathbf{P}_{CP}$ = determinacy coefficient for the correlation preserving predictor; $\mathbf{P}^c_{CP} = \mathbf{P}_{CP}$, corrected for sampling error.

Table S6. Mean determinacy coefficients and bias of parameter-based determinacy coefficients based on Bayes estimation for small and large salient loadings (*sl*), small and large samples (*n*), for variables with different numbers of categories (*c*) for models with zero cross-loadings and zero factor inter-correlations

| sl | n | c | $Cor(\hat{\xi}_{BL},\xi_u)$ | $\mathbf{P}_{BL}$ /$\Delta\mathbf{P}_{BL}$ | $\mathbf{P}^c_{BL}$ /$\Delta\mathbf{P}^c_{BL}$ | $Cor(\hat{\xi}_{CP},\xi_u)$ | $\mathbf{P}_{CP}$ /$\Delta\mathbf{P}_{CP}$ | $\mathbf{P}^c_{CP}$ /$\Delta\mathbf{P}^c_{CP}$ |
|---|---|---|---|---|---|---|---|---|
| .40 | 300 | 2 | **.62 (.02)** | .66 (.03) /.04 | .61 (.04) /-.01 | **.62 (.02)** | .66 (.03) /.04 | .61 (.04) /-.01 |
|  |  | 4 | **.71 (.02)** | .74 (.02) /.03 | .71 (.02) /.00 | **.71 (.02)** | .74 (.02) /.03 | .71 (.02) /.00 |
|  |  | 6 | **.72 (.02)** | .75 (.02) /.03 | .72 (.02) /.00 | **.72 (.02)** | .75 (.02) /.03 | .72 (.02) /.00 |
|  |  | 8 | **.73 (.01)** | .75 (.01) /.02 | .72 (.02) /-.01 | **.73 (.01)** | .75 (.01) /.02 | .72 (.02) /-.01 |
|  | 900 | 2 | **.65 (.01)** | .66 (.01) /.01 | .64 (.01) /.00 | **.65 (.01)** | .66 (.01) /.01 | .64 (.01) /-.01 |
|  |  | 4 | **.72 (.01)** | .73 (.01) /.01 | .72 (.01) /.00 | **.72 (.01)** | .73 (.01) /.01 | .72 (.01) /.00 |
|  |  | 6 | **.73 (.01)** | .74 (.01) /.01 | .73 (.01) /.00 | **.73 (.01)** | .74 (.01) /.01 | .73 (.01) /.00 |
|  |  | 8 | **.74 (.01)** | .75 (.01) /.01 | .74 (.01) /.00 | **.74 (.01)** | .75 (.01) /.01 | .74 (.01) /.00 |
| .80 | 300 | 2 | **.88 (.01)** | .92 (.00) /.04 | .91 (.00) /.04 | **.88 (.01)** | .92 (.00) /.04 | .91 (.00) /.04 |
|  |  | 4 | **.94 (.00)** | .95 (.00) /.01 | .94 (.00) /.00 | **.94 (.00)** | .95 (.00) /.01 | .94 (.00) /.00 |
|  |  | 6 | **.95 (.00)** | .95 (.00) /.00 | .95 (.00) /.00 | **.95 (.00)** | .95 (.00) /.00 | .95 (.00) /.00 |
|  |  | 8 | **.95 (.00)** | .96 (.00) /.00 | .95 (.00) /.00 | **.95 (.00)** | .96 (.00) /.00 | .95 (.00) /.00 |
|  | 900 | 2 | **.88 (.00)** | .92 (.00) /.04 | .92 (.00) /.04 | **.88 (.00)** | .92 (.00) /.04 | .92 (.00) /.04 |
|  |  | 4 | **.94 (.00)** | .95 (.00) /.00 | .95 (.00) /.00 | **.94 (.00)** | .95 (.00) /.00 | .95 (.00) /.00 |
|  |  | 6 | **.95 (.00)** | .95 (.00) /.00 | .95 (.00) /.00 | **.95 (.00)** | .95 (.00) /.00 | .95 (.00) /.00 |
|  |  | 8 | **.95 (.00)** | .96 (.00) /.00 | .95 (.00) /.00 | **.95 (.00)** | .96 (.00) /.00 | .95 (.00) /.00 |

*Note.* Standard deviations are given in brackets; bias of parameter-based determinacy coefficients is marked with "Δ"; score-based determinacy coefficients and bias greater or equal .05 are given in bold face; $Cor(\hat{\xi}_{BL},\xi_u)$ = score-based determinacy coefficient based on the best-linear factor score predictor; $\mathbf{P}_{BL}$ = determinacy coefficient for the best linear predictor; $\mathbf{P}^c_{BL} = \mathbf{P}_{BL}$, corrected for sampling error; $Cor(\hat{\xi}_{CP},\xi_u)$ = score-based determinacy coefficient based on the correlation-preserving factor score predictor; $\mathbf{P}_{CP}$ = determinacy coefficient for the correlation preserving predictor; $\mathbf{P}^c_{CP} = \mathbf{P}_{CP}$, corrected for sampling error.